# Low energy nuclear reactions driven by discrete breathers

V.I. Dubinko
NSC Kharkov Institute of Physics and Technology, Kharkov 61108, Ukraine

A new mechanism of LENR in solids is proposed, which is based on the large amplitude anharmonic lattice vibrations, a.k.a. intrinsic localized modes or "discrete breathers" (DBs). In particular, so called gap DBs, which can arise in diatomic crystals such as metal hydrides, are argued to be the LENR catalyzers. The large mass difference between H or D and the metal atoms provides a gap in phonon spectrum, in which DBs can be excited in the H/D sub-lattice resulting in extreme dynamic closing of adjacent H/D atoms (~ 0.01 Å) required for the tunneling through nuclear Coulomb barrier. DBs have been shown to arise either via thermal activation at elevated temperatures or via knocking atoms out of equilibrium positions under non-equilibrium gas loading conditions, employed under radiolysis or plasma deposition methods. The DB statistics in both cases is analyzed, and an attempt is made to quantify part of the vibrational problem in terms of electrochemical current or ion flux, connecting them with external excitation of DBs that act as *nano-colliders of deuterons* triggering LENR. Resulting analytical expressions (under selected set of material parameters) describe quantitatively the observed exponential dependence on temperature and linear dependence on the electric (or ion) current. Possible ways of engineering the nuclear active environment based on the present concept are discussed.

## 1  Introduction

Numerous experimental data on low energy nuclear reactions assisted by the crystalline environment [1-3] leave little doubts about the reality of LENR, but a comprehensive theory of this phenomenon remains a subject of debates [2-4]. Some of the proposed models attempted to modify conventional nuclear physics by introducing various types of transient quasi-particles and structures such as Hydrino, Hydron, Hydrex etc. that were expected to lower the Coulomb barrier. Other, less radical, models pointed out at the possibility of screening of the Coulomb barrier by atomic electrons. Comprehensive review can be found in refs [2-4]. However, none of these models can explain even qualitatively all salient conditions required for the LENR, which have been summarized by McKubre et al [2] as follows.

LENR observed in heavy water electrolysis at palladium cathodes requires simultaneous attainment of four conditions: (i) high loading of D within the Pd lattice; (ii) an initiation time at least ten times larger than the D diffusion time constant: (iii) a threshold electrochemical surface current or current density that is not correlated to the bulk D loading; (iv) deuterium flux plays an important role in determining the excess heat power density. The first two requirements can be understood as thermodynamic preconditioning needed to bring D atoms within Pd lattice as close as possible. But the most "mysterious" from the point of view of current models are *triggering mechanisms* (iii and iv), which do not depend on the *loading mechanisms* (i and ii). Up to date one can cite the conclusion by McKubre et al [2]: "It is not at all clear how effectively the electron charge transfer reaction or the adsorption/desorption reaction couple energy into modes of lattice vibration appropriate to stimulate D + D interaction."

The present paper is aimed, fist of all, at answering this question. Its main argument is that in crystals with sufficient anharmonicity, a special kind of lattice vibrations, namely, *discrete breathers* (DBs), a.k.a. intrinsic localized modes, can be excited either thermally or by external triggering, in which the amplitude of atomic oscillations greatly exceeds that of harmonic oscillations (phonons) [5-15]. Due to the crystal anharmonicity, the frequency of atomic oscillations increase or decrease with increasing amplitude so that the DB frequency lies outside the phonon frequency band, which explains the weak DB coupling with phonons and, consequently, their *robustness* even at elevated temperatures. DBs have been successfully observed experimentally in various physical systems [8] and materials ranging from metals to diatomic insulators [9], and they have been proposed recently as *catalyzers for various chemical reactions* in solids [16-18]. This filed of research is principally new, and it lies at the conjunction of nonlinear physics with material science. The main message of the present paper is that DBs present a viable *catalyzing mechanism for the nuclear reactions* in solids as well due to the possibility of extreme dynamic closing of adjacent atoms required for the tunneling through the Coulomb barrier.

The paper is organized as follows. In the next section, a short review of the DB properties in metals and diatomic crystals is presented based on results of molecular dynamics (MD) simulations using realistic many-body interatomic potentials. In section 3, a rate theory of DB excitation under thermal heating and under non-equilibrium gas loading conditions (radiolysis or plasma deposition) is developed, and the average rate of D-D "collisions" in DBs is evaluated. In section 4, tunneling through Coulomb barrier with account of electron screening is discussed, and the tunneling coefficient is presented as a function of the barrier width. This is the only section, where conventional *nuclear physics* is employed, which is shown to be just one link in the chain of mechanisms required for the realization of LENR. In section 5, combining the tunneling coefficient with the rate of D-D "collisions" in DBs, the LENR energy production rate is evaluated as a function of temperature, ion (electric) current and material parameters and compared with experimental data. The results are discussed in section 6 and summarized in section 7.

## 2 Discrete breathers in metals and diatomic crystals

Discrete breathers are spatially localized large-amplitude vibrational modes in lattices that exhibit strong anharmonicity [5-8]. They have been identified as exact solutions to a number of model nonlinear systems possessing translational symmetry [8] and successfully observed experimentally in various physical systems [8, 9]. Presently the interest of researchers has shifted to the study of the role of DBs in solid state physics and their impact on the physical properties of materials [9, 16-20].

Until recently the evidence for the DB existence provided by direct atomistic simulations, e.g. molecular dynamics (MD), was restricted mainly to one and two-dimensional networks of coupled nonlinear oscillators employing oversimplified pairwise interatomic potentials [6-8]. Studies of the DBs in three-dimensional systems by means of MD simulations using realistic interatomic potentials include ionic crystals with NaCl structure [10], graphene [12], carbon nanotubes [13] and metals [14, 15, 20].

### 2.1 Metals

For a long time, it has been assumed that the softening of atomic bonds with increasing vibrational amplitude is a general property of crystals, which means that the oscillation frequency decreases with increasing amplitude. Therefore DBs with frequencies above the top phonon frequency were unexpected. However, in 2011, Haas et al [14] have provided a new insight into this problem by demonstrating that the anharmonicity of metals appears to be very different from that of insulators. The point is that the essential contribution to the *screening* of the atomic interactions in metals comes from *free electrons* at the Fermi surface. As a consequence, the ion-ion attractive force may acquire a nonmonotonic dependence on the atomic distance and may be enhanced resulting in an amplification of even anharmonicities for the resulting two-body potentials. This effect can counteract the underlying softening associated with the bare potentials with a moderate increase of vibrational amplitudes to permit the existence of DBs above the top of the phonon spectrum. MD simulations of lattice excitation in fcc nickel as well as in bcc niobium and iron using realistic many-body interatomic potentials have proven that stable high-frequency DBs do exist in these metals [14, 15]. Notably, the excitation energy of DBs can be relatively small (fractions of eV) as compared to the formation energy of a stable Frenkel pair in those metals (several eV).

Moreover, it has been shown that DBs in Fe (and most likely in other transition metals) are highly mobile, hence can efficiently transfer a concentrated vibrational energy over large distances along close-packed crystallographic directions [15, 20]. Recently, a theoretical background has been proposed to ascribe the interaction of moving DBs (a.k.a 'quodons' – quasi-particles propagating along close-packed crystallographic directions[1]) with defects in metals to explain the anomalously accelerated chemical reactions in metals subjected to irradiation. Irradiation may cause continuous generation of DBs inside materials due to *external lattice excitation*, thus 'pumping' a material with DB gas [18, 19].

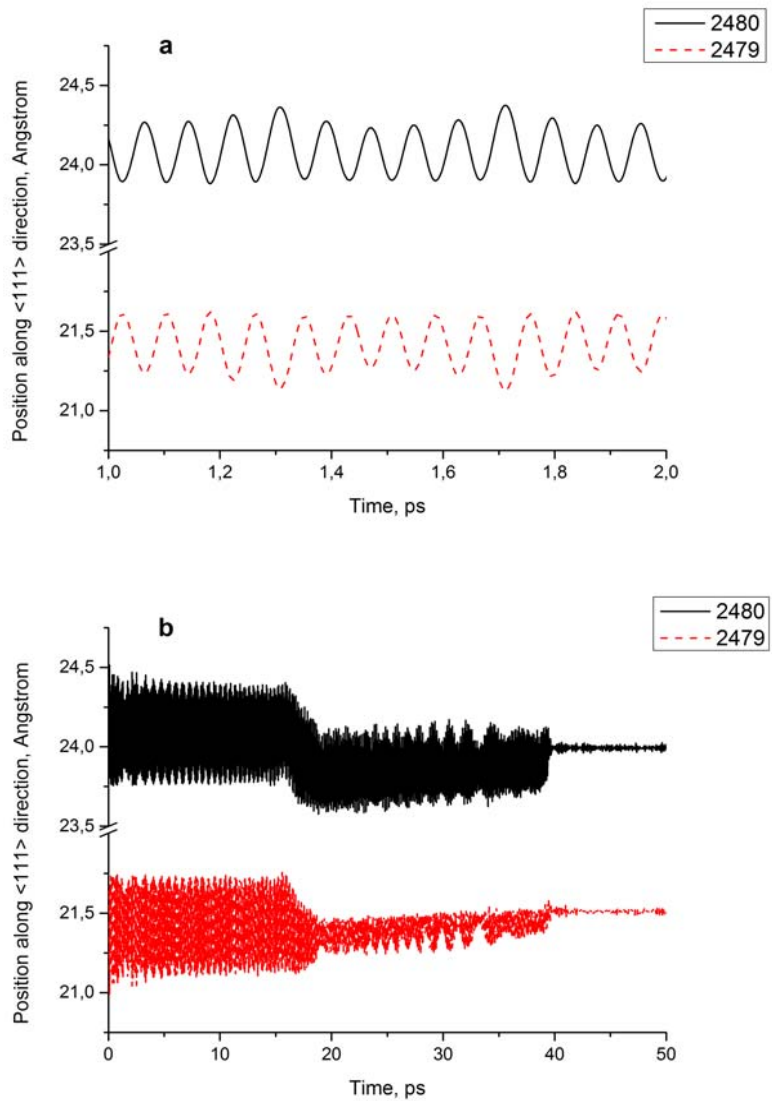

Fig. 1 Oscillation of x coordinate of two neighboring atoms, 2480 and 2479 in a [111] row in Fe in a "standing" DB excited with $d_0$ = 0.325 Å [20]. (a) Initial stage of DB evolution; (b) total lifespan of DB showing a stepwise quantum nature of its decay.

---

[1] It should be noted that Russell and Eilbeck [21] have presented experimental evidence for the existence of quodons that propagate great distances in atomic-chain directions in crystals of muscovite, an insulating solid with a layered crystal structure. Specifically, when a crystal of muscovite was bombarded with alpha-particles at a given point at 300 K, atoms were ejected from remote points on another face of the crystal, lying in atomic chain directions at more than $10^7$ unit cells distance from the site of bombardment.

In order to understand better the structure and properties of standing and moving DBs, consider the ways of their external excitation in Fe by MD simulations [20]. The key feature of the procedure is the initial displacement of the two adjacent atoms from their equilibrium position along the close <111> direction, which should oscillate in the *anti-phase mode* with respect to each other thus forming a stable DB, as shown in Fig. 1 (a). The initial offset displacement $d_0$ determines the DB amplitude and oscillation frequency and, ultimately, its lifetime. DBs can be excited in a narrow frequency band $(1 \div 1.4) \times 10^{13}$ s$^{-1}$ just above the Debye frequency of bcc Fe, and DB frequency grows with increasing amplitude as expected from the "hard" type anharmonicity of the considered vibrational mode. Application of a displacement larger than 0.45 Å generates a chain of *focusons*, while a displacement smaller than 0.27 Å does not provide enough potential energy for the two oscillators to initiate a stable DB and the atomic oscillations decay quickly by losing its energy to *phonons*. The most stable DBs can survive up to 400 oscillations, as shown in Fig. 1 (b), and ultimately decay in a stepwise quantum nature by generating bursts of phonons, as has been predicted by Hizhnyakov as early as in 1996 [22].

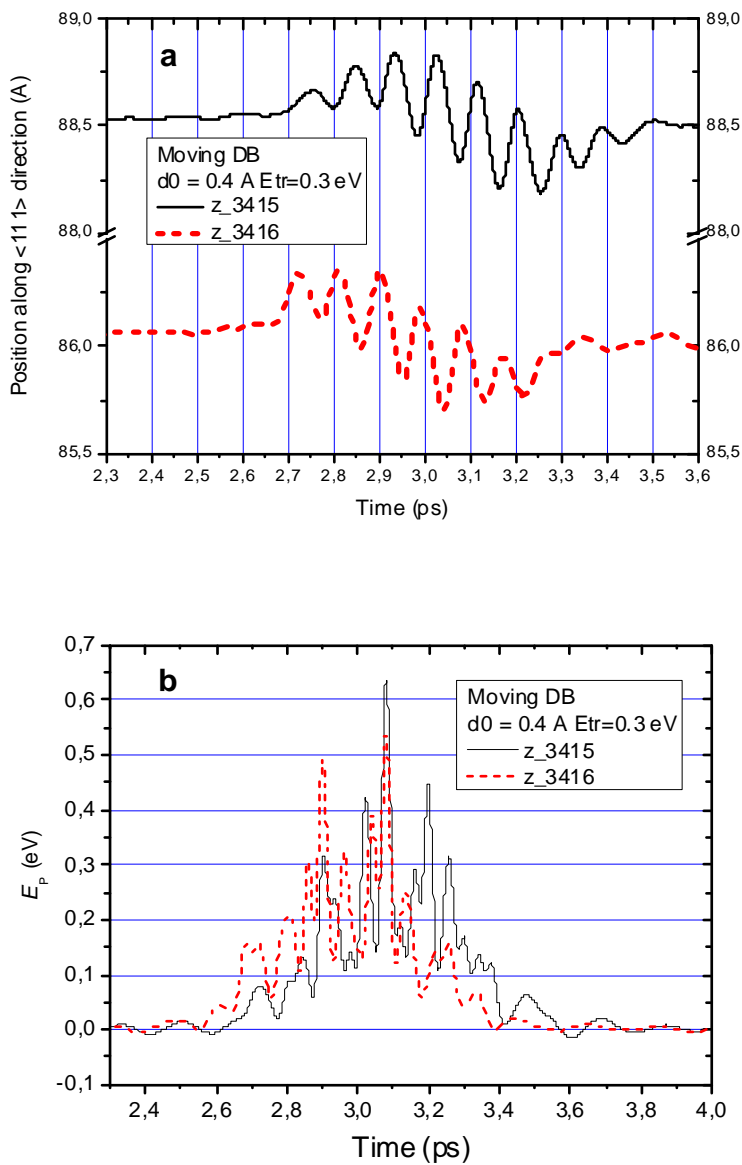

Figure 2 (a) Oscillation of x coordinate of two neighbouring atoms, 3415 and 3416 in a [111] row in Fe during the passage of a moving DB ($d_0 = 0.4$ Å, $E_{tr} = 0.3$ eV); (b) deviation of the potential energy of the atoms from the ground state during the passage of DB [20].

The movement of a DB can be induced by translational kinetic energy $E_{tr}$ given to the two central DB atoms in the same direction along the x-axis. Their velocities range from about 300 to 2000 m/s while travel distances range from several dozens to several hundreds of atomic spaces, depending on the $d_0$ and $E_{tr}$ [20]. Fig.2 (a) shows a DB approaching the atoms with index 3415 and 3416. The two atoms pulsate in the *anti-phase mode* for about 1ps (~10 oscillations) and then oscillations cease but they are resumed at the subsequent atoms along the x-axis. In this way, the DB moves at a speed of 2.14 km/s, i.e. about the half speed of sound in bcc Fe. The translational kinetic energy of the DB is about 0.54 eV, which is shared among two core atoms, giving 0.27 eV per atom. This is very close to the initial kinetic energy $E_{tr}$=0.3 eV transmitted to the atoms to initiate the DB movement. The deviation of the potential energy of the atoms from the ground state during the passage of the DB is presented in Fig.2b. The amplitude of the energy deviation can reach almost 1 eV. In an oversimplified 'thermodynamic' analogy, a moving DB can be viewed as an atom-size spot heated above 1000 K propagating though the crystal at sub-sonic speed.

DB excitations in ref [20] were done in a cell with initial and boundary conditions imitating a perfect crystal at the lattice temperature of 0 K, i.e. when all other atoms were initially at their lattice positions and had zero initial velocities. This poses an important question of the effect of lattice temperature and the crystal size on the robustness of DBs, since the most successful LENR experiments were conducted at temperatures above 300 K and employed the metal (such as Pd, Pt or Ni) based small (even nanosized) particles. Recently, Zhang and Douglas [23] investigated interfacial dynamics of Ni nanoparticles at elevated temperatures exceeding 1000 K and discovered a string-like collective motion of surface atoms with energies in the eV range, i.e. exceeding the average lattice temperature by an order of magnitude. One of the most intriguing observations of this study was the propagation of the *breather excitations* along the strings, providing a possible mechanism for driving the correlated string-like atomic displacement movements. The authors conclude that these dynamic structures might be of crucial significance in *relation to catalysis*.

Now, various electrolytic reactions, which can proceed during the course of absorption/desorption at the cathode surface (see e.g. [4] p. 664 and discussion in the present paper), can be a source of the vibrational energy required for the formation of DBs in the subsurface layer. It should be noted that, based on the MD modeling results, only a fraction of eV may be sufficient to initiate a moving DB along any of the 12 close-packed directions <110> in fcc Pd sub-lattice.

In heavily deuterated palladium, a compound PdD forms, where each octahedral site of the fcc-Pd structure is occupied by one D atom (thus forming another fcc sub-lattice: Fig. 3) and where the lattice constant expands by 6%. Moving DBs may be produced in the D sub-lattice as well. The nearest-neighbor separation between D atoms in PdD is 2.9 Å, while the equilibrium distance between two D atoms inside one octahedral site (in the hypothetical crystal $PdD_2$) is 0.94 Å, as demonstrated by the *ab initio* density-functional calculations [24]. Thus, the equilibrium distance between two D atoms is increased by 0.2 A from the gas value of 0.74 Å, which makes the LENR extremely improbable for any *equilibrium configuration* of D atoms in the palladium lattice. However, a dynamic closing of two D atoms performing *anti-phase oscillations* in a DB within a deuterium sub-lattice cannot be excluded, as will be demonstrated in the following subsection.

*2.2 Diatomic crystals*

Oscillation frequencies of the DBs in ionic crystals are found to be in the gaps of the phonon spectrum, being essentially dependent on *long-range forces*. It means that disregarding the long-range polarization effects, which cause a strong broadening of the optical phonon band, results in totally unrealistic value of the DB frequency in the optical band of the real phonon spectrum [25]. So only a correct account of the localized anharmonic forces with long-range harmonic ones results in adequate description of DBs.

PdD has the NaCl structure, which consists of two face-centered cubic lattices with lattice parameter *a*, one occupied by the light and another one by the heavy ions, displaced with respect to each other by the vector (a/2,0,0) as shown in Fig. 3. In order to preserve its identity, DB must have oscillation frequency and its entire higher harmonics lying outside the phonon bands of the crystal lattice. MD simulations have revealed that diatomic crystals with Morse interatomic interactions typically demonstrate *soft type* of anharmonicity [10], which means that DB's frequency decreases with increasing amplitude, and one can expect to find only so-called gap DBs with frequency within the phonon gap of the crystal. A necessary condition of existence of such DBs is the presence of a sufficiently *wide gap* in the phonon spectrum of the crystal, which can be expected in crystals with components having sufficiently different atomic weights. The atomic weight ratio effect on the properties of DBs was studied by Khadeeva and Dmitriev [10] who used interatomic parameters that are not related to any particular crystal but give stable NaCl structure and realistic values of interaction energies, lattice parameter, vibrational frequencies, etc. This makes their results relevant also for the PdD case even though the explored weight ratio was in the range of 1 to 0.1, i.e. considerably larger than the D/Pd weight ratio of 0.019.

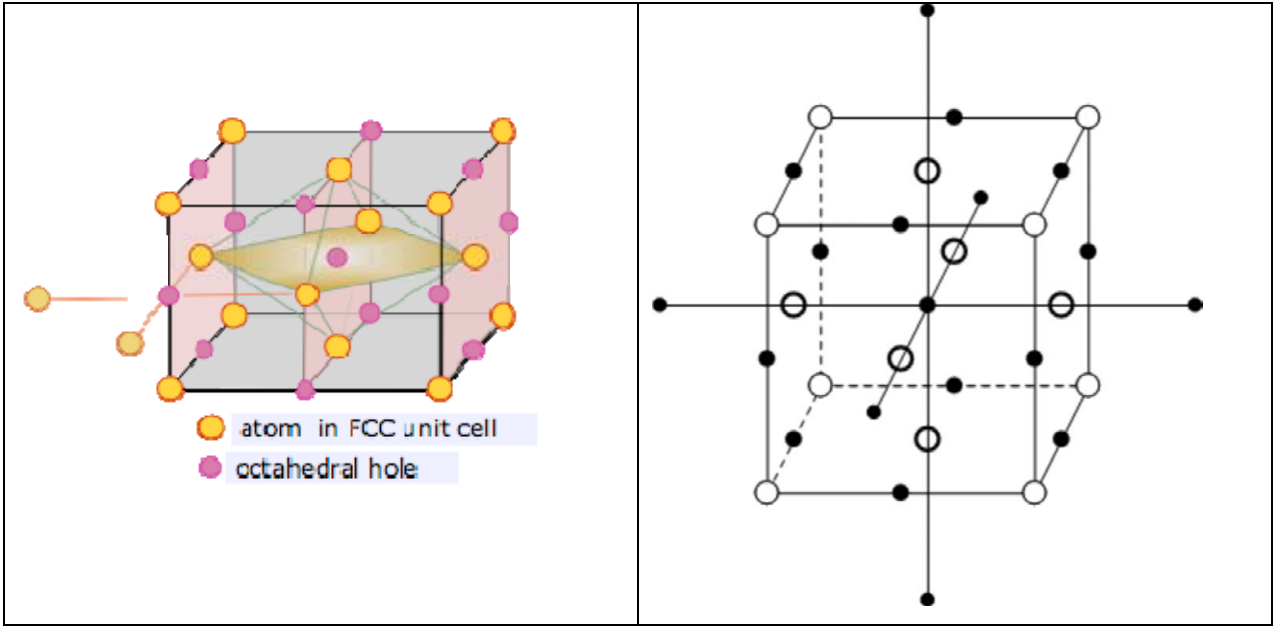

Figure 3. Left: PdD crystal structure. Right: the crystal with NaCl structure modeled in ref [10]. Heavy (light) atoms are shown by open (filled) circles. Adapted from [10] Copyright by APS.

Fig. 4(a) clearly shows that the gap width grows with decreasing weight ratio $M_L/M_K$, and so it might be expected to be wider in the PdD case. The DBs were excited in [10] simply by shifting one *light atom* or two neighboring light atoms from their equilibrium positions while all other atoms were initially at their lattice positions and had zero initial velocities, similar to the DB excitation procedure employed in [20] for Fe. Breather's initial amplitude, $d_0$, was taken from the range $0.01a < d_0 < 0.05a$, where $a = 6.25$ Å is the equilibrium lattice parameter of the NaCl structure. In this way, for the minimal weight ratio $M_L/M_K = 0.1$, three types of stable DBs have been excited, frequencies of which are shown in Fig. 4(b) as the functions of their amplitudes. One can see that the maximum DB amplitude was about 0.35 Å (similar to the iron case [20]), which could bring two adjacent light atoms with initial spacing $b = \sqrt{2}a/2 \approx 1.77$ Å, as close as to the distance of ~1 Å, which, although being much smaller than any phonon-induced closing, is not yet sufficient for the tunneling. However, numerical results on the gap DB in NaI and KI crystals has shown that DB amplitudes along <111> directions can be as high as 1 Å, and the lifetimes can be as long as $10^{-8}$s (more than 20000 oscillations) [6]. Besides, decreasing the weight ratio down to the PdD value of 0.019 can be expected to decrease the minimum attainable spacing within a DB down to a fraction of angstrom, but such MD modeling has not been attempted so far.

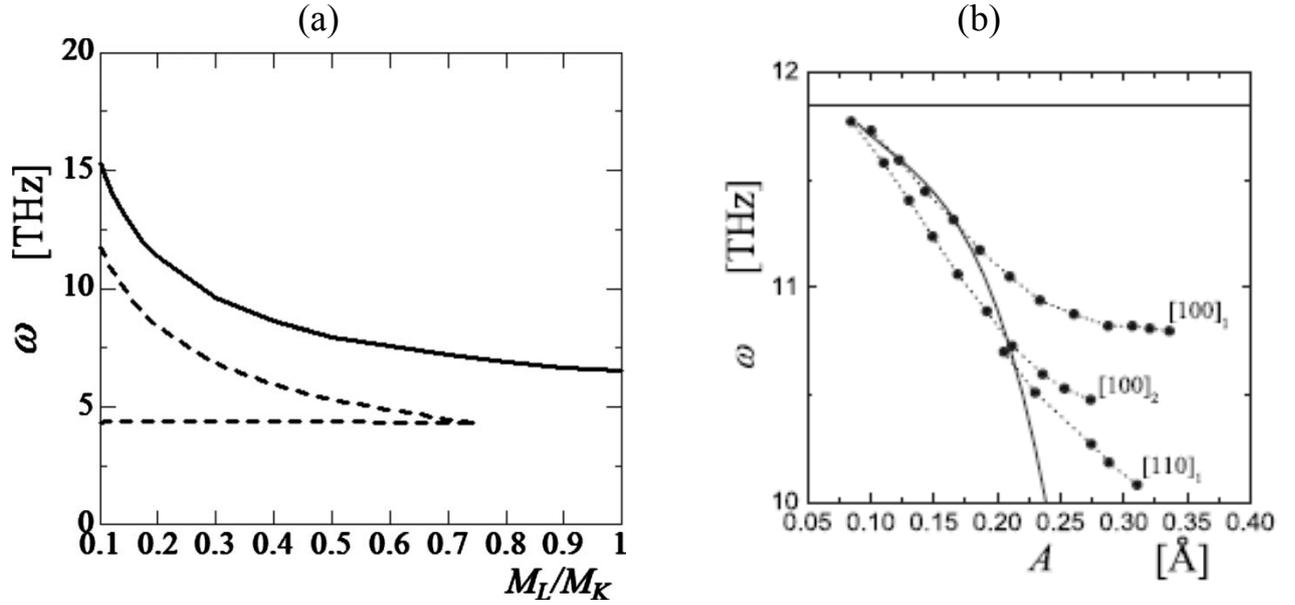

Figure 4. (a) The maximum frequency ω (solid line) and edges of the gap (dashed line) of the phonon spectrum, as the functions of the mass ratio, $M_L/M_K$. (b) Frequencies, as the functions of the DB amplitudes, A, for the DBs of three types: $[110]_1$, $[100]_1$, and $[100]_2$, excited by simulations [10] for $M_L/M_K = 0.1$, where figures in brackets describe polarization and the subscript indicates the number of the atoms oscillating with large amplitude. Solid line gives ω(A) found for the DB $[100]_1$ in frame of the single degree of freedom model developed in [10]. Horizontal line gives the upper edge of the phonon gap. Adapted from [10] Copyright by APS.

Another principle question concerns the mechanisms of excitation and the properties of gap DBs at elevated temperatures. This problem was studied by Kistanov and Dmitriev [26] for the different weight ratios and temperatures. Density of phonon states (DOS) of the NaCl-type crystal for the weight ratio $M_L/M_K = 0.1$ at temperatures ranging from 0 K to 620 K is shown in Fig. 5 (a-d). A small decrease in the gap width of the phonon spectrum with increasing temperature is observed. The appearance of two additional broad peaks in the DOS at elevated temperatures (starting from T = 310 K) is observed. One of them is arranged in the gap of the phonon spectrum, while another one lies above the phonon spectrum. The appearance of the peak *in the gap* of the phonon spectrum can be associated with the spontaneous excitation of gap DBs at sufficiently high temperatures, when nonlinear terms in the expansion of interatomic forces near the equilibrium atomic sites acquire a noticeable role. In connection with this, it has been concluded that as the temperature increases, if the difference in weights of anions and cations is sufficiently large, the lifetime and concentration of gap DBs in the crystal with the NaCl structure increase. The appearance of the peak *above* the phonon spectrum at sufficiently high temperatures can be associated with the excitation of DBs of another type, which manifest the *hard nonlinearity*.

Fig. 5(f) shows DOS for PdD and PdH based on the force constants obtained from the Born von Karman model [27], which quantitatively reproduces the intensity distribution in the experimental S(Q,E) spectrum of PdD obtained in ref. [28] at deuterium pressure of 5 GPa and T=600 K. A slight discrepancy between the calculated and experimental spectra of the second and higher optical bands in PdD suggests a certain *anharmonicity of D vibrations* in these bands [28]. Quantitative estimates [28] show the anharmonicity of the potential well for D atoms at energies as low as ~0.1 eV, counting from the bottom of the well.

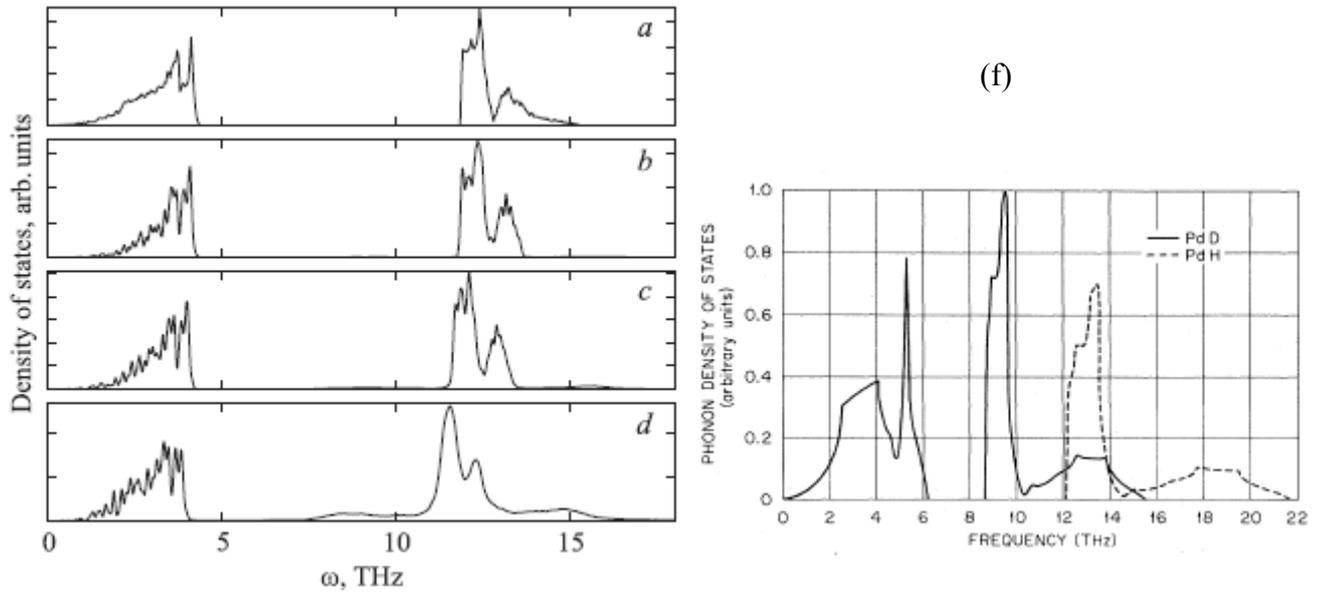

Figure 5. DOS of the NaCl-type crystal for the weight ratio $M_L/M_K = 0.1$ at temperatures T = (a) 0, (b) 155, (c) 310, and (d) 620 K. Adapted from [26] Copyright by APS. (f) DOS for PdD and PdH crystals based on the force constants obtained from the Born von Karman model [27], which quantitatively reproduces the intensity distribution in the experimental S(Q,E) spectrum of PdD obtained in ref. [28] at deuterium pressure of 5 GPa and T=600 K. Adapted from [27] Copyright by APS.

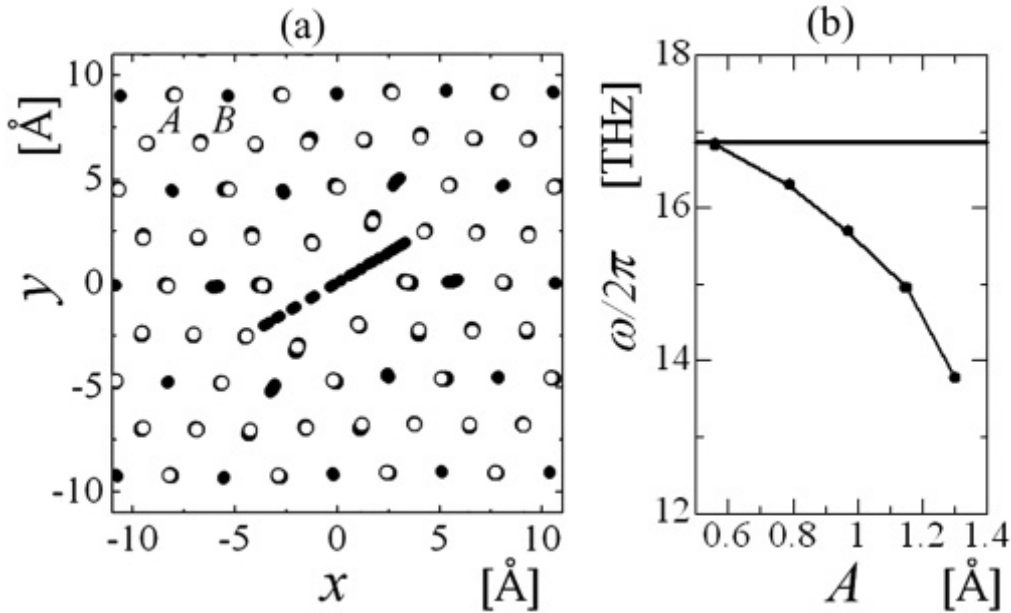

Figure 6. (a) Stroboscopic picture of atomic displacements in the vicinity of gap DB in the crystal $A_3B$ with atomic weight ratio $M_B/M_A = 0.1$ at zero Kelvin. Open (solid) circles correspond to heavy "A" (light "B") atoms. Displacements of atoms are multiplied by a factor of 4. (b) DB frequency as the function of its amplitude. The horizontal line indicates the upper edge of the phonon gap. Adapted from [11] Copyright by APS.

High-temperature excitation of DBs has been demonstrated also for the two-dimensional crystal of $A_3B$ composition with long-range Morse interactions [11]. The lifetime of high-energy atoms was measured for various temperatures for two atomic weight ratios, $M_A/M_B = 0.1$; 0.46. In the first case, the crystal supported gap DBs in the sub-lattice of light atoms, whereas in the other case there were no DBs in the crystal due to the absence of the gap in the phonon spectrum. Fig. 6 shows atomic displacements and DB frequency as the function of its amplitude, which can be as large as 1.3 Å for one oscillating atom at zero Kelvin. It means that two adjacent atoms oscillating in the anti-phase mode could span 2.6 Å (which equals the lattice parameter of the $A_3B$ crystal) and bring the two atoms really close to each other. Excitation of such DBs has not been attempted, but the most important result of ref. [11] is the demonstration of the "natural" thermally-activated way of the DB excitation in a crystal under thermal equilibrium conditions. To do so the authors obtained a crystal at thermal equilibrium using a special thermalization procedure for 100 ps. After that the analysis of *thermal fluctuations* in the crystal was carried out within 200 ps. The temperature of the crystal was characterized by spatially (over the ensemble) and temporally averaged kinetic energy per atom, $\bar{K}$, as $T = \bar{K}/k_B$, where $k_B$ is the Boltzmann constant. Atoms with kinetic energy $K_{A,n}$, $K_{B,n} > e\bar{K}$ were considered to be high-energy atoms, and their energies averaged over the lifetime were evaluated separately for heavy and light atoms. An example of the time evolution of the relative kinetic energy of a particular light atom, $K_{B,n}/\bar{K}$, is presented in Fig. 7 in the units of the DB's oscillation period $\Theta$ for $\bar{K}=0.1$ eV corresponding to the lattice temperature of 1160 K.

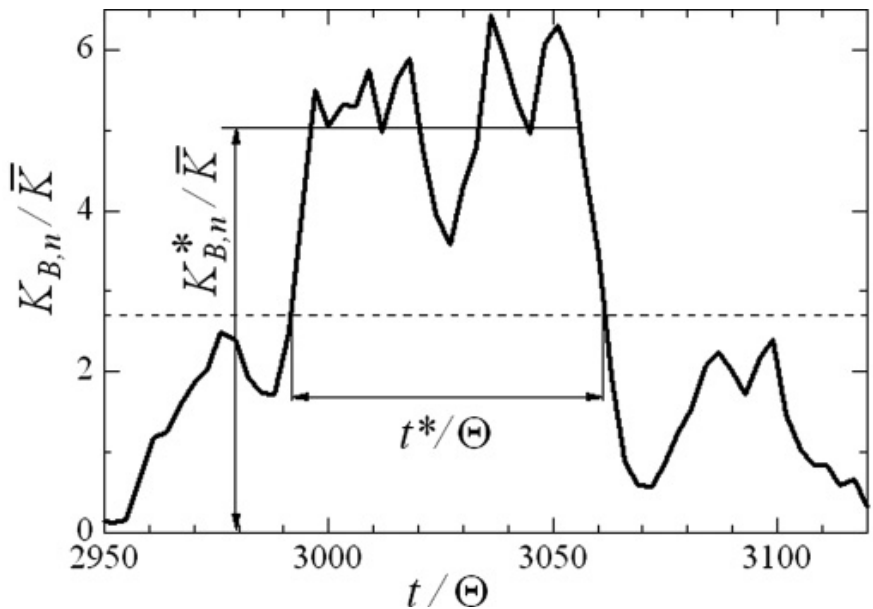

Figure 7. Relative kinetic energy of particular light atom, $K_{B,n}/\bar{K}$, as a function of dimensionless time, $t/\Theta$, where $\Theta$ = 0.06 ps is the DB's period. The kinetic energy of the light atom, $K_{B,n}$, was averaged over time 0.18 ps ≈ 3$\Theta$. The horizontal dashed line indicates the level of kinetic energy equal to $e\bar{K}$, where e ≈ 2.7 is the base of the natural logarithm. The lifetime of the high-energy state of the atom is $t*/\Theta \approx 70$, and its kinetic energy averaged over the lifetime is $K^*_{B,n}/\bar{K} \approx 5.1$. Results are for $M_B/M_A$ = 0.1 and $\bar{K}$ = 0.1 eV. Adapted from [11] Copyright by APS.

The kinetic energy of the light atom, $K_{B,n}$, was averaged over time 0.18 ps ≈ 3$\Theta$. In this example, the lifetime of the high-energy state of an atom is $t*/\Theta \approx 70$, and its kinetic energy averaged over the lifetime is $K^*_{B,n}/\bar{K} \approx 5.1$. Analogously, high-energy heavy atoms were analyzed and concentrations of heavy and light high-energy atoms were evaluated. It appears that the average lifetime and concentration of *high-energy light atoms* increases exponentially with increasing average energy, $\bar{K}$, in a marked contrast with the heavy atoms, which is consistent with the fact that only light atoms have large vibrational amplitudes in DBs. It is the amplitude rather than energy of atomic vibrations that plays a key role in the triggering of LENR, since the probability of tunneling through the Coulomb barrier depends crucially on the separation of D ions before the tunneling (see section 4), which can be drastically reduced in DBs due to the *specific combination of localized (anharmonic) and long-range (harmonic) forces*.

These findings are of primary importance for the concept of LENR driven by DBs, since they point out at the *two ways* of creation of the so called "nuclear active environment" (NAE, as defined by Storms [3]), which is associated with an *environment supporting DBs* in the present paper. The first way is *thermal activation* of DBs in the sub-lattice of D or H within the compound nanocrystal, in which the heavy component is represented by a suitable metal such as Pd, Pt, or Ni. This way seems to be the basic mechanism for the LENR observed e.g. in specially treated nickel surface exposed to hydrogen at high temperatures (see refs.76-79 in [3]). The second way is the DB excitation by *external triggering* such as the atomic displacements in the course of exothermic electrolysis at metal cathodes (majority of LENR experiments) or due to energetic ions, obtained by discharge in gas containing hydrogen isotopes (see refs. 48, 49 in [3]). Naturally, both mechanisms may operate simultaneously under LENR conditions, and this synergy should be reflected in a viable model of DB excitation, the construction of which is attempted in the next section.

## 3  Rate theory of DB excitation under thermal equilibrium and external driving

The rate equation for the concentration of DBs with energy E, $C_{DB}(E,t)$ can be written as follows [17]

$$\frac{\partial C_{DB}(E,t)}{\partial t} = K_{DB}(E) - \frac{C_{DB}(E,t)}{\tau_{DB}(E)}, \tag{1}$$

where $K_B(E)$ is the rate of creation of DBs with energy $E > E_{\min}$ and $\tau_{DB}(E)$ is the DB lifetime. It has an obvious steady-state solution ($\partial C_{DB}(E,t)/\partial t = 0$):

$$C_{DB}(E) = K_{DB}(E)\tau_{DB}(E), \tag{2}$$

In the following sections we will consider the breather formation by thermal activation and then extend the model to non-equilibrium systems with external driving.

### *3.1  Thermal activation*

The exponential dependence of the concentration of high-energy light atoms on temperature in the MD simulations [11] gives evidence in favor of their thermal activation at a rate given by a typical Arrhenius law [7]

$$K_{DB}(E,T) = \omega_{DB} \exp\left(-\frac{E}{k_B T}\right), \tag{3}$$

where $\omega_{DB}$ is the attempt frequency that should be close to the DB frequency. The breather lifetime has been proposed in [7] to be determined by a phenomenological law based on fairly general principles: (i) DBs in two and three dimensions have a minimum energy $E_{\min}$, (ii) The lifetime of a breather grows with its energy as $\tau_{DB} = \tau_{DB}^0 \left(\frac{E}{E_{\min}} - 1\right)^z$, with $z$ and $\tau_B^0$ being constants, whence it follows that under thermal equilibrium, the DB energy distribution function $C_{DB}(E,T)$ and the mean number of breathers per site $n_{DB}(T)$ are given by

$$C_{DB}(E,T) = \omega_{DB}\tau_{DB} \exp\left(-\frac{E}{k_B T}\right), \tag{4}$$

$$n_{DB}(T) = \int_{E_{min}}^{E_{max}} C_{DB}(E,T)dE = \omega_{DB}\tau_{DB}^{0} \frac{\exp\left(-\frac{E_{min}}{k_BT}\right)}{(E_{min}/k_BT)^{z+1}} \int_{0}^{\frac{E_{max}-E_{min}}{k_BT}} y^z \exp(-y)dy, \qquad (5)$$

Noting that $\Gamma(z+1,x) = \int_{0}^{x} y^z \exp(-y)dy$ is the second incomplete gamma function, eq. (5) can be written as [17]

$$n_{DB} = \omega_{DB}\tau_{DB}^{0} \frac{\exp(-E_{min}/k_BT)}{(E_{min}/k_BT)^{z+1}} \Gamma\left(z+1, \frac{E_{max}-E_{min}}{k_BT}\right), \qquad (6)$$

It can be seen that the mean DB energy is higher than the averaged energy density (or temperature):

$$\langle E_{DB} \rangle = \frac{\int_{E_{min}}^{E_{max}} C_{DB}(E,T)E\,dE}{\int_{E_{min}}^{E_{max}} C_{DB}(E,T)dE} \xrightarrow{E_{max} \gg E_{min}} \left(\frac{E_{min}}{k_BT} + z + 1\right) \times k_bT, \qquad (7)$$

Assuming, according to [11] (Fig. 5), that $E_{min}/k_BT \approx 3$ and $\langle E_B \rangle \approx 5k_BT$, one obtains an estimate for $z \approx 1$, which corresponds to linear increase of the DB lifetime with energy.

*3.2 External driving*

Fluctuation activated nature of DB creation can be described in the framework of classical Kramers model, which is archetypal for investigations in reaction-rate theory [29]. The model considers a Brownian particle moving in a symmetric double-well potential $U(x)$ (Fig. 8(a)). The particle is subject to fluctuational forces that are, for example, induced by coupling to a heat bath. The fluctuational forces cause transitions between the neighboring potential wells with a rate given by the famous Kramers rate:

$$\dot{R}_K(E_0,T) = \omega_0 \exp(-E_0/k_BT), \qquad (8)$$

where $\omega_0$ is the attempt frequency and $E_0$ is the height of the potential barrier separating the two stable states, which, in the case of fluctuational DB creation, corresponds to the minimum energy that should be transferred to particular atoms in order to initiate a stable DB. Thus, the DB creation rate (3) is given by the Kramers rate: $K_{DB}(E,T) = \dot{R}_K(E,T)$.

In the presence of *periodic modulation* (driving) of the well depth (or the reaction barrier height) such as $U(x,t) = U(x) - V(x/x_m)\cos(\Omega t)$, the reaction rate $\dot{R}_K$ averaged over times exceeding the modulation period has been shown to increase according to the following equation [17]:

$$\langle \dot{R} \rangle_m = \dot{R}_K I_0\left(\frac{V}{k_bT}\right), \qquad (9)$$

where the amplification factor $I_0(x)$ is the zero order modified Bessel function of the first kind. Note that the amplification factor is determined by the ratio of the modulation amplitude $V$ to temperature, and it does not depend on the modulation frequency or the mean barrier height. Thus, although the periodic forcing may be too weak to induce *athermal* reaction (if $V < E_0$), it can amplify the average reaction rate drastically if the ratio $V/k_BT$ is high enough, as it is demonstrated in Fig. 8(b).

Another mechanism of enhancing the DB creation rate is based on small *stochastic modulations* of the DB activation barriers caused by external driving. Stochastic driving has been shown to enhance the reaction rates via effective reduction of the underlying reaction barriers [18, 19] as:

$$\langle \dot{R} \rangle = \omega_0 \exp\left(-E_a^{DB}/k_b T\right), \quad E_a^{DB} = E_0 - \frac{\langle V \rangle_{SD}^2}{2k_b T}, \tag{10}$$

where $\langle V \rangle_{SD}$ is the standard deviation of the potential energy of atoms surrounding the activation site.

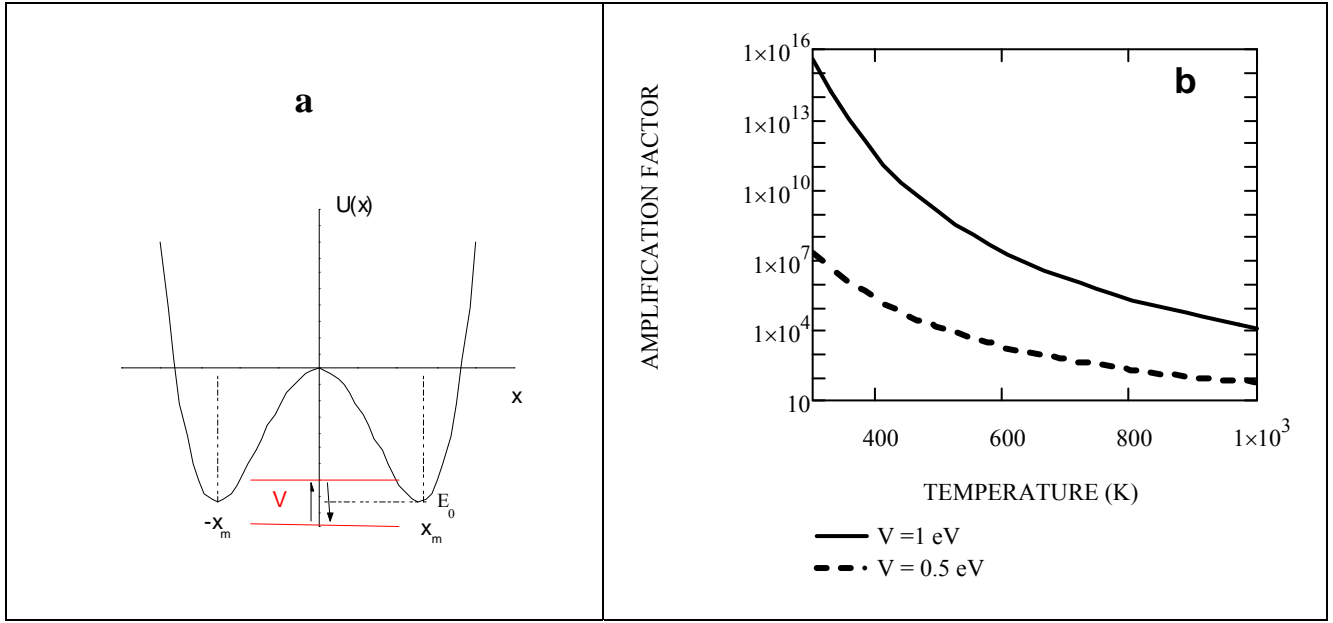

Figure 8. (a) Sketch of the double-well potential landscape with minima located at $\pm x_m$. These are stable states before and after reaction, separated by a potential "barrier" with the height changing periodically or stochastically within the V band. (b) Amplification factor, $I_0(V/k_B T)$, for the average escape rate of a thermalized Brownian particle from a periodically modulated potential barrier at different temperatures and modulation amplitudes, V [17].

Consider the periodic driving of the DB creation in more details. It can be provided, e.g. by focusons or *quodons* formed by knocking of surface atoms out of equilibrium position by energetic ions or molecules under non-equilibrium deposition of deuterium, which may be sufficient to initiate a quodon propagating inside the material along a close-packed direction up to the depth equal to the quodon propagation range, $l_q$. The amplitude of the quasi-periodic energy deviation of metal atoms along the quodon pathway can reach almost 1 eV with the excitation time, $\tau_{ex}$, of about 10 oscillation periods (Fig. 2). In the modified Kramers model (9), this energy deviation corresponds to the modulation of the DB activation barrier. Then, a *macroscopic* rate of generation of DBs of energy $E_{DB}$ (per atom per second) may be written as follows:

$$\langle K_{DB} \rangle_{macro} = \omega_{DB} \exp\left(-\frac{E_{DB}}{k_B T}\right)\left(1 + \left\langle I_0\left(\frac{V_{ex}}{k_b T}\right)\right\rangle \omega_{ex} \tau_{ex}\right), \tag{11}$$

where $V_{ex}$ is the excitation energy and $\omega_{ex}$ is the mean number of excitations per atom per second caused by the flux of quodons, $F_q$, which is proportional to the electric current density, $J$, in the case of radiolysis (or to the ion flux, for the plasma deposition) and is given by

$$\omega_{ex}(F_q) = F_q b^2 \frac{4\pi R_P^2 l_q}{4\pi R_P^3/3} = F_q b^2 \frac{3l_q}{R_P}, \quad F_q(J) = \frac{1}{2} J \tag{12}$$

where $b$ is the atomic spacing, and so the product $F_q b^2$ is the frequency of the excitations per atom within the layer of the quodon propagation of a thickness $l_q$, while the ratio $3l_q/R_P$ is the geometrical factor that corresponds to the relative number of atoms within the quodon range in a PdD particle of a radius $R_P$. The coefficient of proportionality between $F_q$ and the electron flux $J/e$ (where $e$ is the electron charge) assumes that each electrolytic reaction that involves a pair of electrons, releases a vibrational energy of ~1 eV, which is sufficient for generation of one quodon with energy $V_q \approx V_{ex} < 1$ eV. Multiplying the DB generation rate (11) by their lifetime $\tau_{DB}$ and the oscillation frequency $\omega_{DB}$ one obtains the mean frequency of D-D "collisions" in DBs (per atom per second) as a function of temperature, electric current density and material parameters listed in Table 1:

$$W_{D-D}(T,J) = (\omega_{DB})^2 \tau_{DB} \exp\left(-\frac{E_{DB}}{k_B T}\right)\left(1 + \left\langle I_0\left(\frac{V_{ex}}{k_b T}\right)\right\rangle \omega_{ex}(J)\tau_{ex}\right), \qquad (13)$$

Fig. 9 shows that the DB-induced mean frequency of D-D "collisions" grows linearly with electric current density and exponentially with temperature, which results in a deviation from the linear dependence of $W_{D-D}$ on $J$, if temperature increases with increasing electric current density.

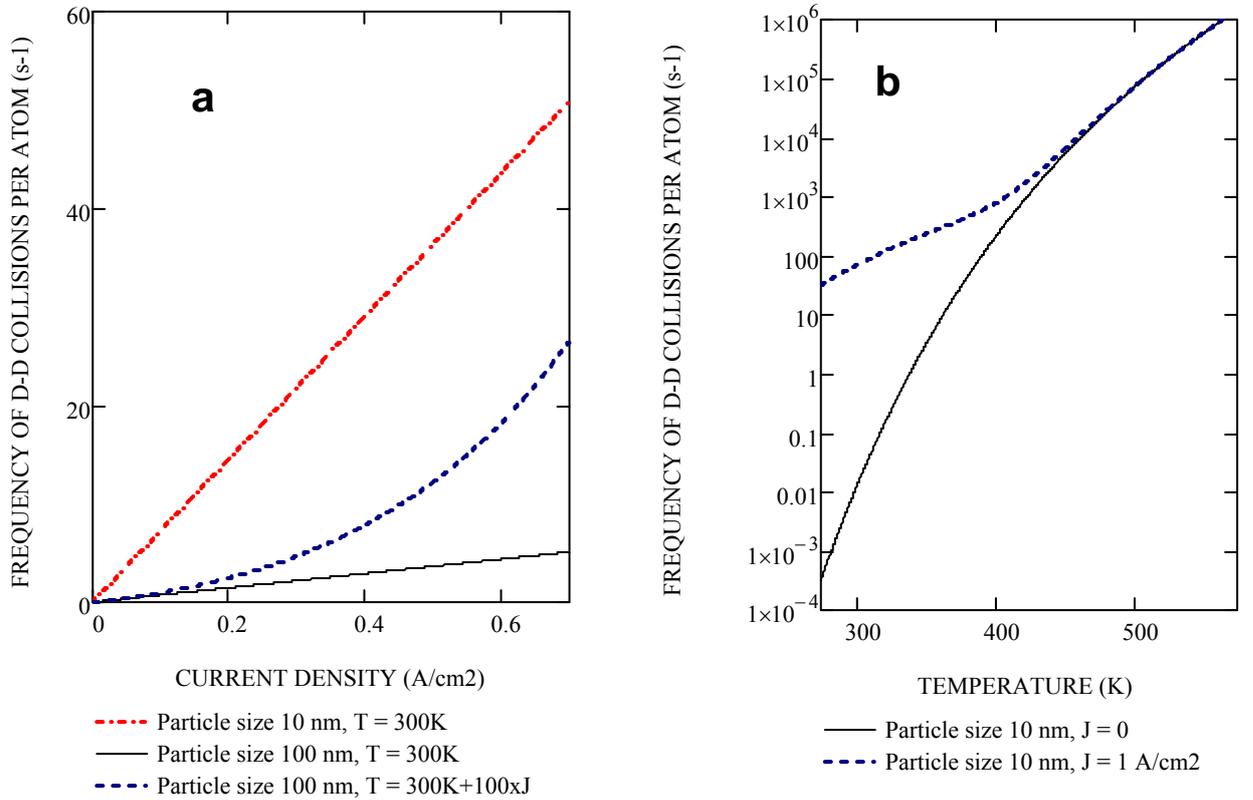

Figure 9. Mean frequency of D-D "collisions" in DBs as a function of electric current density, temperature assuming material parameters listed in Table 1

In this way, DBs can provide up to $10^{25}$ "collisions" per cubic cm per second, which can be regarded as attempts to penetrate the Coulomb barrier and initiate LENR, provided that the minimum D-D spacing attainable within DBs is sufficiently small. In the next section we estimate the required spacing by evaluating the tunneling coefficient as a function of the barrier width with account of electron screening.

## 4 Tunneling with account of electron screening

The tunneling coefficient (TC) first derived by Gamow (1928) for a pure Coulomb barrier is the Gamow factor, given by

$$G \approx \exp\left\{-\frac{2}{\hbar}\int_{r_1}^{r_2} dr \sqrt{2\mu(V(r)-E)}\right\} \quad (14)$$

Where $2\pi\hbar$ is the Plank's constant, $E$ is the nucleus CM energy, $\mu$ is the reduced mass, $r_1$, $r_2$ are the two classical turning points for the potential barrier, which for the D-D reaction are given simply by $\mu = m_D/2$, $V(r) = e^2/r$. For two D's at room temperature with thermal energies of $E \sim 0.025$ eV, one has $G \sim 10^{-2760}$, which explains a pessimism about LENR and shows a need for possible mechanisms of electron shielding of the barrier in a solid.

We will use a conventional model of a spherical shell of radius R of negative charge surrounding each D, which results essentially in a shifted Coulomb potential

$$V(r) = e^2\left[\frac{1}{r} - \frac{1}{R}\right], \quad r_n \leq r \leq R \quad (15)$$

where $r_n \ll R$ is the nuclear well radius. In this model, one has a modified TC ([4] p. 620)

$$G^*(R) = \exp\left\{-\frac{2\pi e^2}{\hbar}\sqrt{\frac{\mu}{2(E+e^2/R)}}\right\} \quad (16)$$

which depends crucially on the D ions spacing before the tunneling, as shown in Fig. 10.

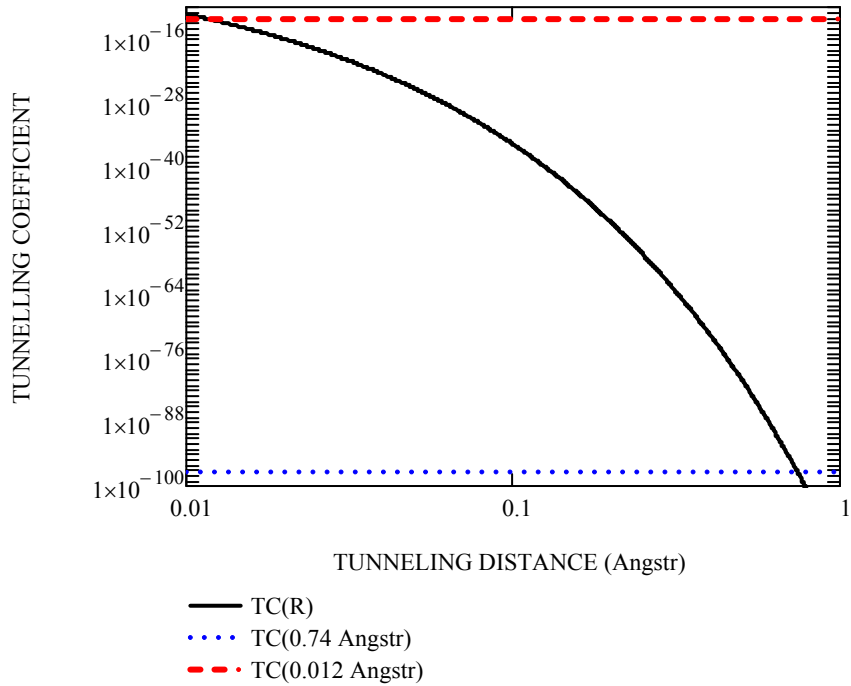

Figure 10. Tunneling coefficient dependence on the tunneling distance R for two deuterons with thermal energies of $E \sim 0.025$ eV. The horizontal lines correspond to R=0.74 Å (equilibrium D-D spacing in $D_2$ molecule) and to R=0.012 Å (minimum dynamic spacing required for LENR by the DB mechanism).

This model is the simplest both conceptually and computationally, and so it cannot pretend at giving one the *exact* values, which can differ from the model values by many orders of magnitude. But this model demonstrates the most important property of the electron screening, namely, the *equivalence of the deuteron kinetic energy needed for the tunneling and the screening depth*, R: they enter into denominator of eq. (16) as a sum $E + e^2/R$, which means that decreasing screening depth R increases the tunneling probability as strongly as increasing deuteron's energy up to the level of $e^2/R$. It can be seen that in the angstrom range expected in any equilibrium D-D structure, the tunneling probability is too low for practical applications, but it picks up ~100 orders of magnitude with R decreasing down to 0.01 Å. This R range is lower by almost two orders of magnitude than the range of *conventional chemical forces*, but it is higher by three orders of magnitude than the range of *nuclear forces*, and the main hypothesis of the present paper is that it can be attainable in DBs due to the *specific combination of localized (anharmonic) and long-range (harmonic) forces.* Based on this hypothesis, the LENR energy production rate is evaluated in the next section as a function of temperature and electric current density and compared with experimental data.

It should be noted that the reaction rate per deuteron-deuteron pair evaluated in the next section, based on eq (16), does not take into account many physical effects, such as the volume factor and nuclear potential, which could results in corrections to the reaction rate of 5-6 orders of magnitude. However, one can see that changing the screening depth by only 0.01 Å results in the change of the tunneling probability by 5 orders of magnitude, and so making the model more complex does not seem to make it more accurate.

## 5 LENR energy production rate under heavy water electrolysis

We consider the following reaction [2]
$$D + D \rightarrow {}^4He + 23.8\ MeV_{lattice} \qquad (17)$$
which is based on experimentally observed production of excess heat correlated with production of a "nuclear ash", i.e. $^4$He [2, 3]. Multiplying the DB-induced mean frequency of D-D "collisions" (13) by the tunneling probability in each collision (16) and the energy $E_{D-D}$=23.8 MeV, produced in D-D fusion one obtains the LENR power production rate per atom, $P_{D-D}$:
$$P_{D-D}(T,J) = W_{D-D}(T,J) G^*(R_{DB}) E_{D-D}, \qquad (18)$$
Usually, experimentalists measure the output power density per unite surface of a macroscopic cell, $P_{D-D}^S$, as a function of the electric current density, as demonstrated by many researchers ( see e.g. [3] p. 77) and illustrated in Fig. 11. This is given by the product of $P_{D-D}$, the number of atoms per unite volume, $1/\omega_{PdD}$ ($\omega_{PdD}$ being the atomic volume of PdD) and the ratio of the cell volume to the cell surface:
$$P_{D-D}^S(T,J) = \frac{P_{D-D}(T,J)}{\omega_{PdD}} \frac{L_S^3}{L_S^2} = P_{D-D}(T,J) \frac{L_S}{\omega_{PdD}}, \qquad (19)$$
where $L_S$ is the cell size, if cubic, or thickness, in case of a plate.

Fig. 12 shows the LENR output power density DBs as a function of electric current density and temperature evaluated by eq. (19) assuming material parameters listed in Table 1. Comparison of Figs. 12 (a) with experimental data (Fig. 11) shows that the present model describes quantitatively the observed linear dependence of $P_{D-D}^S$ on the current density at a constant temperature as well as the deviation from the linear dependence, if temperature increases with increasing electric current density. Thermally-activated nature of the reactions leading to LENR has been noted for quite a long time (see e.g. [3] p. 76: In general, the larger the temperature, the more excess energy is reported), and the activation energy was estimated in some cases to be ~15 kcal/mol $\approx$ 0.65 eV. The present model not only explains these observations, but reveals that the underlying physics is a consequence of the synergy between thermally-activated and externally-driven mechanisms of the DB excitation in deuterated palladium, which results in a violation of the classical Arrhenius law and in the renormalization of the underlying activation energies [17].

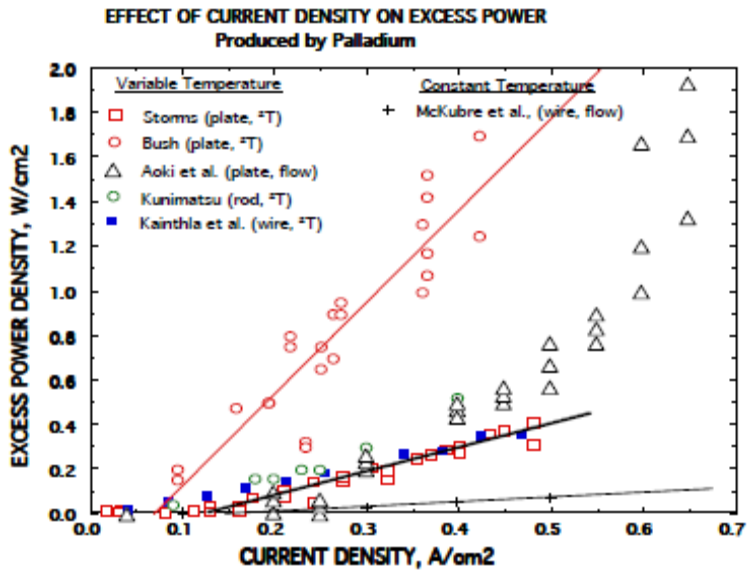

Figure 11. Comparison between several studies showing the effect of current density on power density [3].

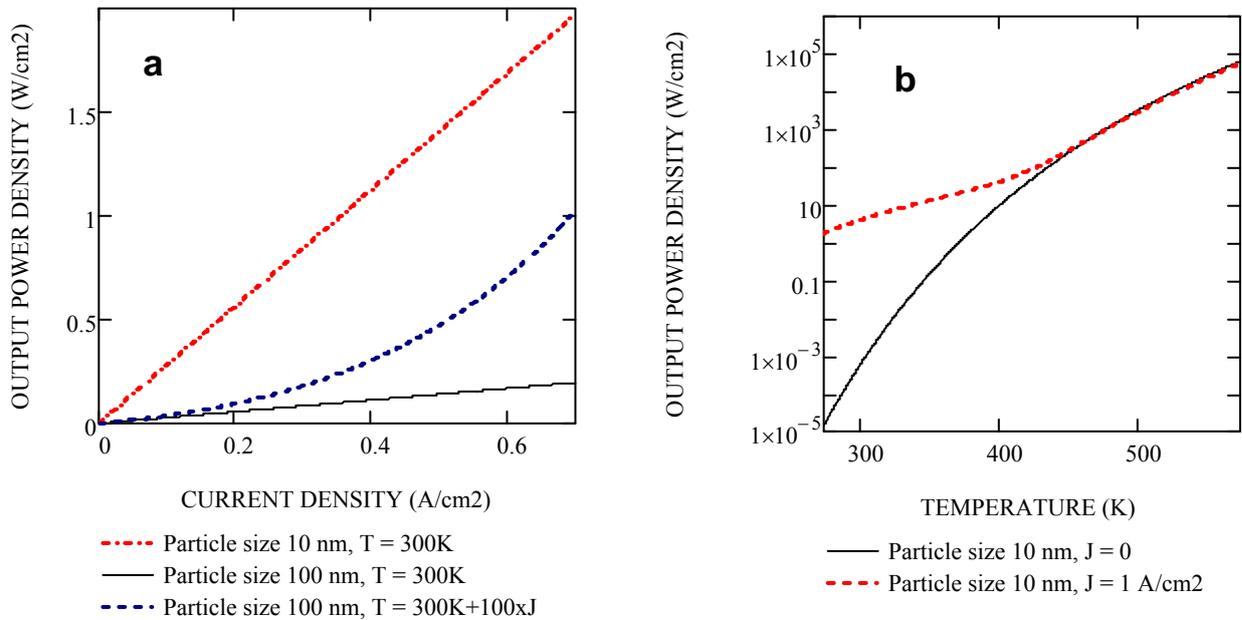

Figure 12. LENR output power density as a function of electric current density and temperature evaluated by eq. (19) assuming material parameters listed in Table 1.

It should be noted that among the material parameters listed in Table 1 the most important is the minimum attainable D-D spacing in DBs, $R_{DB}$, since its increase from the fitted value of 0.012 Å to 0.1 Å suppresses the LENR rate by 23 order of magnitude (Fig. 13) which could not be compensated by any choice of other material parameters within the framework of the present model.

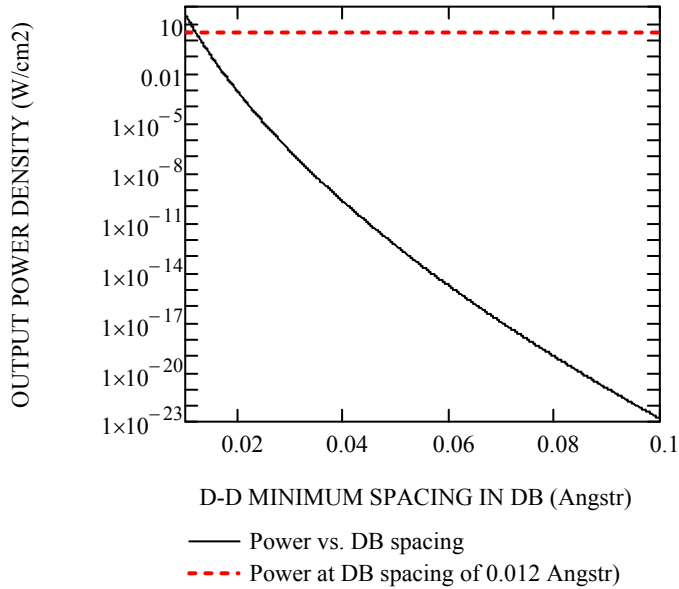

Figure 13. LENR output power density as a function of D-D minimum spacing in DBs evaluated by eq. (19) at $J = 1$ A/cm$^2$, $T - 300$ K assuming material parameters listed in Table 1.

## 6 Discussion

In the introduction we sited the problem formulated by McKubre et al [2] concerning the coupling of the adsorption/desorption reaction energy into modes of lattice vibration appropriate to stimulate D + D interaction. Indeed, in spite of a number of models trying to take into account the phonons, i.e. packets of wave-energy present in a lattice, as the LENR drivers (see e.g. refs to Hagelstein, Swartz, and F. S. Liu in [3]), one could not help feeling that something important was missing in the theory. Phonons were expected to move energy between nuclei, thereby creating enough localized energy to overcome the Coulomb barrier. But phonons are plane harmonic waves, essentially *delocalized* in the crystal, and the amplitude of atomic vibrations in harmonic range does not exceed ~0.1 Å, which is absolutely insufficient for bringing atoms closely enough for any significant tunneling (Fig. 13), whatever the underlying mathematics is. In contrast to phonons, DBs, also known as *intrinsic localized modes*, are essentially localized atomic vibrations that have large amplitudes of ~ 1 Å, which, at least in principle, can bring atoms very close to each other in the anti-phase oscillation mode. DBs can be excited ether thermally at sufficiently high temperatures (which are above the temperature range of typical radiolysis) or by external triggering producing atomic displacements in the subsurface layer, which facilitate the DB creation.

In this view, the first two LENR requirements: (i, ii) high loading of D within the Pd lattice and long initiation time, can be understood as preconditioning needed to prepare PdD particles of small sizes out of initially bulk Pd crystal, in which DBs can be excited, and the requirements (iii, iv) on the D flux and electric current density, are natural prerequisites for the DB creation by the input energy transformed into the lattice vibrations.

The importance of the Fermi-Pasta-Ulam type of energy localization in nano-PdD particles was stressed also by Ahern [30]: Energy localized vibrational modes are so large that they can break and reform bonds. Locally, the vibrations act like very hot regions with active chemistry. Ahern sustained that the phenomenon of localization of energy could even explain the triggering of LENR in nanostructures saturated with hydrogen/deuterium, which is confirmed by the present model. In this connection, it should be noted that for the first time to our knowledge, the idea to link the nonlinear mechanisms of energy localization in the forms of quodons with the Lattice Assisted Nuclear Fusion (LANF), was expressed by Russell and Eilbeck at the LENCOS-09 conference in 2009, which proceeding were published in 2011 [31]. It was suggested that once a quodon is created then atoms of high speed are repeatedly brought close together in head-on collisions without expenditure of energy. If this happened in a crystal containing deuterium, then there would be an enhanced finite probability for fusion to occur, no matter how small the fusion cross-section might be. The proposed LANF mechanism was based on rather high quodon energies (up to ~ 300 eV) with essentially unlimited lifetimes (and propagation ranges) that could be excited in a LiD *crystal ring* by external driving. Although estimates [31] leaded the authors to conclude that the contribution of quodons to additional fusion was negligible, the expressed ideas gave a strong motivation for further research in this direction to the present author. While more recent MD modeling of DBs in real crystals demonstrated an importance of *low-energy atomic collisions* (~ eV) as drivers of LENR via the DB excitation.

An important implication of the present theory is that it may be helpful in selecting a frequency interval for the terahertz stimulation of DBs in deuterated palladium, similar to the method by Letts and Hagelstein [32] of the stimulation of optical phonons by a laser. The stimulation was provided by tuning dual lasers to one of three specific beat frequencies, which were believed to correspond to the frequencies of optical phonons of deuterated palladium. Indeed, the excess power has been produced peaking at frequencies of 8, 15 and 20 THz. However, the comparison of these frequencies with experimentally verified the DOS for PdD crystals [27, 28], shows that the first frequency corresponded to the *gap in the phonon spectrum* of PdD while the last two frequencies were *above the phonon band*, as demonstrated in Fig. 5 (f). So the results [32] give some evidence in favor of the laser-induced stimulation of the "soft" gap DBs as well as of the "hard" high-frequency ones. It should be noted here that the position of the phonon gap may be questioned due to the uncertainty connected with extent of D loading in the experiment [32], but the existence of the highest resonance peak at ~20 THz present a puzzle for the phonon mechanism of LENR. On the other hand, the appearance of high-frequency (hard-type) breathers in NaCl type crystals at elevated temperatures was discussed and demonstrated recently in [11] (see Fig. 5 (d)). This problem clearly needs further investigations as well as the perspective way of the laser-induced stimulation of DB creation in nuclear (i.e. breather) active environment.

Probably the most serious criticism of the present theory may be due to extremely small D-D separation $R_{DB} \approx 0.012$ Å that should be attainable within DBs in order to provide the tunneling probability required to fit the experimental data. Such distances are larger than those, at which nuclear forces operate, by three orders of magnitude, but they are considerably smaller than the range of conventional chemical forces based on electronic properties. So the question arises whether electrons can be effective in screening the Coulomb barrier at distances small enough in order to significantly facilitate the nuclear tunneling? Recent experiments on fusion of elements in accelerators, analyzed in ref. [33], give a positive answer to this question. The analysis is based on corrections to the cross section of the fusion due to the screening effect of atomic electrons, obtained in so called Born-Oppenheimer static approximation by Assenbaum et al [34]. Using this approach, it was shown that the so-called "screening potential" acts as an additional energy of collision at the center of mass. Apparently, this approach is equivalent to taking into account the *reduced thickness* of the potential barrier for calculation of the tunneling probability adopted in the present paper (eq. (16)). However, other models for the fusion reaction rate should be considered in future, where the tunneling rate could be determined more adequately from the time average deuteron-deuteron overlap.

The screening potential for D-D fusion in platinum for deuterons with kinetic energy ranging from 4 to 23 keV was found to be 675 ± 50 eV, which is 25 times larger than for free atoms of deuterium. This may mean that in a crystal of platinum, deuterons of such energies do not feel the Coulomb repulsion up to distances of 25 times smaller than the size of the deuterium atoms. The screening potential for palladium was reported to be ~300 eV. These results provide experimental evidence in favor of the electron screening as such, but it should be stressed that they do not represent the complicated cooperative dynamics of ions and electrons in a DB. Indeed, the velocity of a deuteron accelerated up to 4 keV is about $10^8$ cm/s, which exceeds the atomic velocities in DBs almost by three orders of magnitude. Such deuteron movement is too fast for effective electron screening, since Fermi electrons move at comparable speed, in contrast to the slow D movement in a DB, which allows electrons to follow ions adiabatically. Thus we may conclude that screening distances under DB conditions may be significantly smaller that those obtained under accelerator conditions, due to *specific combination of localized (anharmonic) and long-range (harmonic) forces* acting on deuterons under dynamic oscillation within a discrete breather. In a crude analogy, one can imagine a DB as a *nano-collider* of deuterons, which operates with energies within the eV range rather than the keV range examined in the 'low energy' beam experiments. There is no analytical theory developed for these conditions, and the only data available so far could be obtained from the MD or *ab initio* modeling. So the minimal D-D dynamic spacing attainable in DBs should be regarded as an *adjustable parameter* of the model (all of which are listed Table 1).

In the existing electrochemical loading models (see e.g. ref [35]), three reactions are used: the Volmer reaction, the Tafel reaction and the Heyrovsky reaction, which complicate a quantitative analysis of the vibrational energy produced during the loading. But it should be emphasized that only a fraction of eV may be sufficient for the DB creation, and the experimentally confirmed presence of vacancies in PdD during electrolysis [34] gives evidence of that such energy is indeed deposited, since vacancies require ~1 eV to form in Pd. What is more, vacancies don't diffuse near room temperature by thermal activation; so the only way they can be formed in the Fleischmann-Pons experiments was thought to be through inadvertent codeposition of Pd at high loading [34]. The present concept offers an alternative explanation of the accelerated formation and diffusion of vacancies in crystals via DB-induced acceleration of chemical reactions, as have been argued in ref. [18]. So the present model simply assumes that each electrolytic reaction that involves a pair of electrons, releases a vibrational energy of ~1 eV, which is sufficient for the generation of one quodon, and the quodon flux accelerates thermally activated DB creation in sub-surface layers.

On the first glance, the DB persistence and robustness even at high temperatures may look astonishing. However, DBs do not radiate their energy in the form of small-amplitude waves because they vibrate at frequencies outside the phonon spectrum of crystal. DB frequency can leave the phonon spectrum when its amplitude is sufficiently large because the frequency of a nonlinear oscillator is amplitude-dependent. There are two types of nonlinearity, the hard-type and the soft-type. In the former (latter) case the DB frequency increases (decreases) with increase in its amplitude. In the case of the hard-type nonlinearity the DB frequency can be above the phonon spectrum. For the soft-type nonlinearity DBs can exist only if the phonon spectrum possesses a gap, which is the case e.g. in the crystals with the NaCl structure with large mass difference of atoms, *which is the case for the PdD crystal*. Numerical results on the gap DB in NaI and KI crystals has shown that DB amplitudes along <111> directions can be as high as 1 Å, and the lifetimes can be as long as $10^{-8}$s (more than 20000 oscillations) [6]. So the mean DB lifetime (100 oscillations) and the excitation time (10 oscillations) assumed in the present model for a quantitative comparison with experiment are rather modest estimates based upon recent MD modeling of DBs using large scale classical MD simulations in 3D periodic bcc Fe [20]. Two well spread interatomic potentials (IAPs) have been checked, derived to account for the electronic charge distribution depending on the local atomic arrangement which are known to provide a good compromise between computationally expensive *ab initio* calculations and over-simplified pairwise potentials. What is more, the latest DFT (*ab initio*) calculations, which are independent from the choice of IAP, have shown that a standing DB in Fe can be stable up to 160 oscillations [36].

One of the fundamental questions concerning the robustness of DBs in metals is connected with phonon-electron relaxation in metals, which is one of the reasons behind the fast optical phonon relaxation in metals. In fact, the electron-phonon coupling is a basic mechanism of the DB creation in *thermal spikes* in metals produced by fast electrons and especially by fast heavy ions (with energy exceeding 1 MeV/amu) that lose their energy via the excitation of electron system rather than via collisions with crystal atoms [17]. Although DBs (similar to lattice phonons) may lose part of their energy to free electrons, they can also gain energy from the phonons. So a DB should be regarded as an open dissipative system rather than an isolated one, which provides a natural explanation of thermally-activated creation of DBs at elevated and high temperatures demonstrated in recent works [11, 23, 26]. Besides, it should be noted that the realistic MD and especially DFT calculations take into account the electronic charge distribution adjusting adiabatically to the local atomic arrangement under DB conditions.

Small size of PdD particles is required since the triggering of DB creation occurs due to the propagation of the vibrational energy from the surface (by quodons, focusons etc.) down to some depth, and the smaller is the particles the more atoms can be involved in the DB creation, i.e. become "nuclear active". This is manifested in the model by the inversely proportional dependence of the power output on the particle size (see expression (12) for excitation frequency $\omega_{ex}$ and Fig. 12 (a)). Storms [3] underlies that "not all small particles are nuclear-active, other factors must play a role as well". From the point of view of the present model, this can be explained by two factors: (i) *impurity atoms* and (ii) *crystal disorder*. Impurity atoms can strongly affect the phonon spectrum of PdD. Although impurity atoms are localized and their concentration may be low, they may change the phonon spectrum of *the whole crystal* and extend it into the DB range, which would suppress the DB formation and make the particle "nuclear inactive" (or vise versa!). This consideration may be a useful tool for the search of the "nuclear active environment" (NAE) by the way of doping the Metal-D or Metal-H crystals with elements changing the phonon spectrum so that to mediate the DB creation.

Another factor concerns the role of the crystal disorder in LENR, which typically start after prolonged loading resulting in formation of numerous defects or occurs in specially prepared fine powders consisting of nanosized particles [2, 3]. As noted by Storms [3], p. 123 : "Cracks and small particles are the Yin and Yang of the cold fusion environment. Small particles are created between cracks and crack-like gaps are formed in the near-contact regions between small particles. The greater the numbers of cracks, the smaller are the isolated regions (particles) between cracks". The present model offers the following hypothesis for the importance of the crystal disorder caused by structural defects and nano-dimensions of the particles.

Piazza and Sanejouand [37] reported a striking *site selectiveness* of energy localization in the presence of spatial disorder. In particular, while studying thermal excitation of DBs in protein clusters, they found that, as a sheer consequence of disorder, *a non-zero energy gap* for exciting a DB at a given site either exists or not.[2] Remarkably, in the former case, the gaps arise as a result of the impossibility of exciting small-amplitude modes in the first place. In contrast, in the latter case, a small subset of linear edge modes acts as accumulation points, whereby DBs can be continued to arbitrary small energies, while unavoidably approaching one of such normal modes. Concerning the structure–dynamics relationship, the authors [37] found that the regions of protein structures where DBs form easily (*zero or small gaps*) were unfailingly the most highly connected ones, also characterized by weak local clustering. This result means that the process of loading or special "nano-treatment" creates the disordered cluster structures, which may be enriched with sites of *zero or small gaps* for the DB excitation. Such cites are expected to become the nuclear active cites, according to the present model. The most important consequence of this hypothesis is that it may offers the ways of *engineering* the nuclear active environment based on the MD modeling of DB creation in nanoparticles and disordered structures.

Finally, it is known that LENR have been shown to produce excess heat and $^4$He as the only significant "nuclear ash", which is hard to explain in the framework of conventional nuclear physics. It has been argued that if a mechanism to overcome the Coulomb barrier is proposed, a mechanism to release the energy must be proposed at the same time, and these two mechanisms must be able to work together. The second mechanism may require modification of the known nuclear reactions, which is beyond the scope of the present paper, and has been discussed extensively, e.g. in a recent review [38]. However, the author believes that the method to overcome the Coulomb barrier proposed in the present paper is more important from a practical point of view, since it may suggest new ways of engineering the NAE by preparing relevant cluster structures so that to facilitate in them creation of discrete breathers as the most constitutive catalyzer of LENR.

---

[2] For a long time it was known that in two-dimensional and three-dimensional *perfect lattices* there is always a *non-zero* energy gap for exciting a DB [7].

## 7 Summary

A new mechanism of LENR in solids is proposed, in which DBs play the role of a catalyzer via extreme dynamic closing of adjacent H/D atoms required for the tunneling through the Coulomb barrier. DBs have been shown to arise either via thermal activation at elevated temperatures or via knocking atoms out of equilibrium positions under non-equilibrium gas loading conditions, employed under radiolysis or plasma deposition methods.

The present mechanism explains all the salient LENR requirements: (i, ii) long initiation time and high loading of D within the Pd lattice as preconditioning needed to prepare small PdD crystals, in which DBs can be excited more easily, and (iii, iv) the triggering by D flux or electric current, which facilitates the DB creation by the input energy transformed into the lattice vibrations.

An attempt is made to quantify part of the vibrational problem in terms of electrochemical current or ion flux and connect it with external triggering of the DB creation subsequently leading to the triggering of LENR. Simple analytical expressions for the cold fusion energy production rate are derived as the functions of temperature, ion (electric) current and material parameters. These expressions (under selected set of material parameters) describe quantitatively the observed exponential dependence on temperature and linear dependence on the electric (or ion) current.

The present results are based only on the *known physical principles* and on independent atomistic simulations of DBs in metals and ion crystals using realistic many-body interatomic potentials. Further research in this direction is needed and planned in order to verify the proposed mechanism by atomistic simulations of DBs in Metal-D and Metal-H systems. An outstanding goal of this research is to suggest new ways of *engineering the nuclear active environment* by preparing relevant cluster structures (by special doping and mechanical treatment) so that to facilitate in them creation of discrete breathers as the most constitutive *catalyzer* of LENR.

**Acknowledgements**

The author is grateful to Mike Russell who attracted his attention to the potential importance of anharmonic lattice vibrations in LENR, to Francesco Piazza for the note on breathers in disordered structures, to Juan Archilla for interesting discussions and to Vladimir Hizhnyakov for valuable criticism and hospitality during a stay at the Tartu University.

**References**

[1] M. Fleischmann, S. Pons, M. Hawkins, Electrochemically induced nuclear fusion of deuterium. *J. Electroanal. Chem.* **261** (1989) 301-308 and errata in Vol. 263, 187-188.
[2] M. McKubre, F. Tanzella, P. Hagelstein, K. Mullican, M. Trevithick, *The Need for Triggering in Cold Fusion Reactions*. in Tenth International Conference on Cold Fusion (MA: LENR-CANR.org., Cambridge, 2003)
[3] Storms, E.K., *The science of low energy nuclear reaction* (World Scientific, Singapore, 2007).
[4] V.A. Chechin, V.A. Tsarev, M. Rabinovich, Y.E. Kim, *Critical review of theoretical models for anomalous effects in deuterated metals*, Int. Journ. of Theoretical Physics, **33** (1994) 617-670.
[5] A.J. Sievers and S. Takeno, *Intrinsic Localized Modes in Anharmonic Crystals*, Phys. Rev. Lett. **61** (1988) 970-973.
[6] V. Hizhnyakov, D. Nevedrov, A. J. Sievers, *Quantum properties of intrinsic localized modes*, Physica B, **316–317** (2002) 132-135.
[7] F. Piazza, S. Lepri, R. Livi, *Cooling nonlinear lattices toward energy localization*, Chaos **13** (2003) 637-645..
[8] S. Flach, A.V. Gorbach, *Discrete breathers — Advances in theory and applications*, Phys. Rep. **467**, (2008) 1-116.
[9] M.E. Manley, *Impact of intrinsic localized modes of atomic motion on materials properties*, Acta Materialia, **58** (2010) 2926-2935.

[10] L.Z. Khadeeva, S.V. Dmitriev, *Discrete breathers in crystals with NaCl structure*, Phys. Rev. B **81** (2010) 214306-1- 214306-8.

[11] L.Z. Khadeeva, S.V. Dmitriev, *Lifetime of gap discrete breathers in diatomic crystals at thermal equilibrium*, Phys. Rev. B, **84** (2011) 144304-1 - 144304-8.

[12] B. Liu, C.D. Reddy, J. Jiang, J.A. Baimova, S.V. Dmitriev, A.A. Nazarov, K. Zhou, *Discrete breathers in hydrogenated graphene*, J. Phys. D-Appl. Phys. **46** (2013) 305302-1-305302-9.

[13] T. Shimada, D. Shirasaki, Y. Kinoshita, Y. Doi, A. Nakatani and T. Kitamura, Infuluence of nonlinear atomic interaction on excitation of intrinsic localized modes in carbon nanotubes, Physica D, **239** (2010) 407-413.

[14] M. Haas, V. Hizhnyakov, A. Shelkan, M. Klopov, *Prediction of high-frequency intrinsic localized modes in Ni and Nb*, Phys. Rev. B, **84** (2011) 144303-1-144303-8.

[15] V. Hizhnyakov, M. Haas, A. Shelkan, M. Klopov, *Theory and molecular dynamics simulations of intrinsic localized modes and defect formation in solids,* Phys. Scr. **89** (2014) 044003 (5pp).

[16] V.I. Dubinko, F.M. Russell, *Radiation damage and recovery due to the interaction of crystal defects with anharmonic lattice excitations*, J. Nuclear Materials, 419 (2011) 378-385.

[17] V. I. Dubinko, P. A. Selyshchev, and J. F. R. Archilla, *Reaction-rate theory with account of the crystal anharmonicity*, Phys. Rev. E **83** (2011) No 4, doi: 10.1103/PhysRevE.83.041124.

[18] V. I. Dubinko, A. V. Dubinko, *Modification of reaction rates under irradiation of crystalline solids: contribution from intrinsic localized modes*, Nuclear Inst. and Methods in Physics Research, B **303** (2013) 133–135.

[19] V. Dubinko, R. Shapovalov, *Theory of a quodon gas. With application to precipitation kinetics in solids under irradiation*. (Springer International Publishing, Switzerland, 2014).

[20] D. Terentyev, A. Dubinko, V. Dubinko, S. Dmitriev, E. Zhurkin, *Interaction of discrete breathers with primary lattice defects in bcc Fe*, Submitted to Europhys. Lett.

[21] F.M. Russell, J.C. Eilbeck, *Evidence for moving breathers in a layered crystal insulator at 300 K,* Europhys. Lett., **78** (2007) 10004-10012.

[22] V. Hizhnyakov, *Relaxation jumps of strong vibration*, Phys. Rev. B, **53** (1996) 13981-13984.

[23] H. Zhang, J. F. Douglas, *Glassy interfacial dynamics of Ni nanoparticles: Part II Discrete breathers as an explanation of two-level energy fluctuations*, Soft Matter, **9** (2013) 1266-1280.

[24] Z. Sun and D. Tomanek, *Cold Fusion: How Close Can Deuterium Atoms Come inside Palladium?*, Phys. Rev. Letters 63 (1989) 59-61.

[25] M. Haas, V. Hizhnyakov, M. Klopov, A. Shelkan, *Effects of long-range forces in nonlinear dynamics of crystals: creation of defects and self-localized vibrations,* in 11th Europhysical Conference on Defects in Insulating Materials (EURODIM 2010) doi:10.1088/1757-899X/15/1/012045.

[26] AA. Kistanov, S.V. Dmitriev, *Spontaneous excitation of discrete breathers in crystals with the NaCl structure at elevated temperatures,* Phys. Solid State, **54** (2012) 1648–1651.

[27] J. M. Rowe, J. J. Rush, H. G. Smith, M. Mostoller, and H. E. Flotow, .*Lattice dynamics of a single crystal of PdD$_{0.63}$,* Phys. Rev. Lett. **33** (1974) 1297-1300.

[28] V.E. Antonov, A.I. Davydov, V.K. Fedotov, A.S. Ivanov., A.I. Kolesnikov, M.A. Kuzovnikov *Neutron spectroscopy of H impurities in PdD: Covibrations of the H and D atoms*, Phys. Rev. B. **80** (2009) 134302-1-134302-7.

[29] Hanggi P, Talkner P, Borkovec M. *Reaction-rate theory: fifty years after Kramers*, Rev. Mod. Phys. **62**, (1990) 251-341.

[30] B. Ahern, "Energy Localization, the key to understanding energy in nanotechnology and nature", http://lenr-canr.org/acrobat/AhernBSenergyloca.pdf

[31] F. M. Russell, J. C. Eilbeck, *Persistent mobile lattice excitations in a crystalline insulator*, In: Discrete and Continuous Dynamical Systems - Series S, Vol. 4, No. 5, 10.2011, p. 1267-1285.

[32] D. Letts, P.L. Hagelstein. *Stimulation of Optical Phonons in Deuterated Palladium*. in ICCF-14 International Conference on Condensed Matter Nuclear Science. (Washington, DC, 2008).

[33] E. N. Tsyganov, *Cold nuclear fusion,* Physics of Atomic Nuclei, **75** (2012) 153–159.
[34] H. J. Assenbaum, K. Langanke C. Rolfs, *Effect of electron screening on low-energy fusion cross section*, Z. Phys. A **327** (1987) 461-468.
[35] S. Chubb, T. Dolan, *Summary of the 2010 Colloquium on Lattice-Assisted Nuclear Reactions at MIT,* http://www.infinite-energy.com/images/pdfs/Colloquium2010.pdf
[36] V.I. Dubinko, D.A. Terentyev, S.V. Dmitriev, V. Hizhnyakov, A. J. Sievers, *Discrete breathers in Iron: ab initio simulations and physical effects*, to be published.
[37] F. Piazza, Y. H. Sanejouand, *Discrete breathers in protein structures*, Phys. Biol. **5** (2008) 026001 (14pp) doi:10.1088/1478-3975/5/2/026001
[38] V.F. Zelensky, *Nuclear processes in deuterium/natural hydrogen - metal systems,* Problems of atomic science and technology, **N3(85)** (2013), Series: Nuclear Physics Investigations (60), p.76-118.

Table 1. Material and DB parameters used in calculations

| Parameter | Value |
| --- | --- |
| D-D equilibrium spacing in PdD, $b$ (Å) | 0.29 |
| D-D minimal dynamic spacing in DB, $R_{DB}$ (Å) | 0.012 |
| DB minimal activation energy, $E_{min}$ (eV) | 0.1 |
| Mean DB energy, $E_{DB}$ (eV) | 1 |
| DB oscillation frequency, $\omega_{DB}$ (THz) | 10 |
| Mean DB lifetime, $\tau_{DB} = 100/\omega_{DB}$ (s) | $10^{-11}$ |
| Quodon excitation energy $V_q \approx V_{ex}$ (eV) | 0.8 |
| Quodon excitation time, $\tau_{ex} = 10/\omega_{DB}$ (s) | $10^{-12}$ |
| Quodon propagation range, $l_q = 10b$ (nm) | 2.9 |
| Cathod size/thickness (mm) | 5 |